\documentclass[10pt,a4paper,twoside]{article}
\usepackage{epsfig}
\usepackage{baltlat5}
\usepackage{wrapfig}
\begin{document}
\ \

\vspace{0.5mm}

\setcounter{page}{1}

\vspace{5mm}

\titlehead{Baltic Astronomy, vol.\ts XX, XXX--XXX, 2005.}

\titleb{SPECTROSCOPIC SEARCH FOR BINARIES\\ AMONG EHB STARS
IN GLOBULAR CLUSTERS~\footnote{\footnotesize
Based on observations with the ESO Very Large
Telescope at Paranal Observatory, Chile (proposal ID 69.D-0682).} }

\begin{authorl}

\authorb{C.~Moni~Bidin}{1,2}
\authorb{R.A.~Mendez}{2}
\authorb{S.~Moehler}{3}
\authorb{G.~Piotto}{1}
\authorb{A.~Recio-Blanco}{1,4}
\authorb{Y.~Momany}{1}

\end{authorl}

\begin{addressl}

\addressb{1}{Dipartimento di Astronomia,
Universtit\'a di Padova, Vicolo dell'osservatorio 2,
35122 Padova, Italy}
\addressb{2}{Departamento de Astronom\'ia, Universidad de Chile,
Casilla 36-D. Santiago, Chile}
\addressb{3}{Institut f\"{u}r Theoretische Physik und Astrophysik,
Christan-Albrechts-Universit\"at zu Kiel, 24098 Kiel, Germany}
\addressb{4}{Observatoire de la C\^{o}te d'Azur, Dpt. Cassiop\'ee, CNRS UMR 6202,
B.P. 4229, 06304 Nice, Cedex 04, France}
\end{addressl}

\submitb{Received 2005 XXX}

\begin{abstract}

We performed a spectroscopic search for binaries among hot Horizontal
Branch stars in globular clusters. We present final results
for a sample of 51 stars in NGC\,6752, and preliminary results
for the first 15 stars analyzed in M\,80.
The observed stars are distributed along all the HBs in the range
8000~$\leq$~T$_{\mathrm{eff}}$~$\leq$~32000~K, and have been observed
during four nights. Radial velocity variations have been measured
with the cross-correlation technique. We carefully analyzed the statistical
and systematic errors associated with the measurements in order to
evaluate the statistical significance of the observed variations.
No close binary system has been detected, neither among cooler stars
nor among the sample
of hot EHB stars (18 stars with T$_{\mathrm{eff}}\geq$~22000~K in NGC\,6752).
The data corrected for instrumental effects
indicate that the radial velocity variations are always below the
3$\sigma$ level of $\approx$~15~km~s$^{-1}$. These results are in sharp
contrast with those found for field hot subdwarfs, and open
new questions about the formation of EHB stars in globular clusters, and
possibly of the field subdwarfs.

\end{abstract}


\resthead{Spectroscopic search for binaries among EHB stars in
Glubular Clusters}{C.~Moni Bidin et al.}

\sectionb{1}{INTRODUCTION}

Although stellar evolution theory has successfully identified
Horizontal Branch (HB) stars as post-core helium flash stars of
low initial mass (Hoyle \& Schwarzschild 1955; Faulkner 1966), we still
lack a comprehensive understanding of their nature. There is a general
agreement that the hottest HB stars (EHB) must have suffered a heavy mass
loss during their evolution, keeping only a thin envelope
($\approx$~0.02~\msun), but their specific formation mechanism remains
unclear. The binarity of EHB stars, as proposed by many authors
(Mengel et al.~1976; Heber et al.~2002) can provide an explanation,
since in a close binary system the mass loss can be enhanced through a
number of different binary evolution channels (Han et al.~2002).
Binaries have been found to be very common among field subdwarf B-type stars
(sdBs) stars, considered to be the counterparts of the cluster EHB stars.
Maxted et al.~(2001) estimated
that 69$\pm$9\% of sdB stars are close binaries with periods P
$\leq$~10~days, and Han et al.~(2003), from binary population synthesis
techniques, predicted a 76-89\% binary fraction. From the recent
investigation of Morales-Rueda et al.~(2003) it appears clear that
close binaries with periods P~$\leq$~5~days and semiamplitudes of the
radial velocity variation K~$\geq$~50~km~s$^{-1}$ are very common
among sdBs. Nevertheless the binary scenarios, extensively
investigated by Han et al.~(2002), are not free from problems as shown
by Lisker et al.~(2005). Moreover a more recent survey of Napiwotzki
et al.~(2004) found a significantly lower binary fraction among
the sdB sample (42\%), showing that we are still far from a full
understanding of this kind of stars. With this contribution we
investigate a sample of EHB stars in globular clusters, searching
for evidence of the presence of close binaries among them.

\sectionb{2}{OBSERVATIONS AND DATA REDUCTION}

\begin{wrapfigure}[15]{l}[0pt]{5.5cm}
\vskip-2mm
\vbox{
\tabcolsep=4pt
\begin{tabular}{c|c c c c}
\multicolumn{5}{c}{\parbox{5cm}{
~~~~{\bf Table 1.}{\
UT of the start of the exposures (hour and minutes).}}}\\
\tablerule
field & \multicolumn{4}{|c}{night} \\
\tablerule
 & \multicolumn{1}{c}{12} & 13 & 14 & 15 \\
\tablerule
A  & 8:43 & 2:44 & 9:34 & 7:47 \\
  & & 3:56 & & \\
\tablerule
B  & 5:55 & 5:08 & 8:33 & 8:58 \\
  & 6:59 & & & \\
\tablerule
C & - - & 8:04 & 6:39 & 6:36\\
  & & & 7:33 & \\
\tablerule
\end{tabular}
}
\end{wrapfigure}
\vskip2mm

In the globular cluster NGC\,6752 we selected 51 target stars from the
photometric data of Momany et al.~(2002). We chose to perform a more complete
investigation distributing the sample along the entire HB,
from cooler stars (T$_{\mathrm{eff}}$~$\approx$~8000~K)
to hot EHB stars (T$_{\mathrm{eff}}\geq$~20000~K), although our attention
was focused toward the EHB sample among which we expected to find a high fraction
of close binary systems.
The targets were divided into three stellar
fields for multiobject spectroscopy.
The position of the targets in the color-magnitude diagram of the
cluster is shown in Figure 1. In Figure 2 we indicate their radial
distribution with respect to the cluster center. The spectra were
collected during four
nights of observation (from 2002 June 12th to 15th) at the VLT-UT4
telescope with the FORS2 spectrograph in MXU mode. During each night,
up to 2 pairs of 1800s exposures were secured in each field, with
grism 1400V+18 (0.5'' wide slits, 1.2~\AA\ resolution), except in the
3rd night in field A, where only one single exposure was
acquired, and the 1st night in field C, where no observations were
performed. In Table 1, the UT of the start of the pair of 1800s
exposures is shown.

Due to different positions of the slits on the mask, the spectra
covered slightly different spectral ranges, but in each spectrum the
H$_{\beta}$ line was always present, except for star 14 in field C,
that has been excluded from our analysis.

\begin{wrapfigure}[41]{r}[0pt]{62mm}
\vskip-1mm
\vbox{
\vskip1mm
\centerline{\psfig{figure=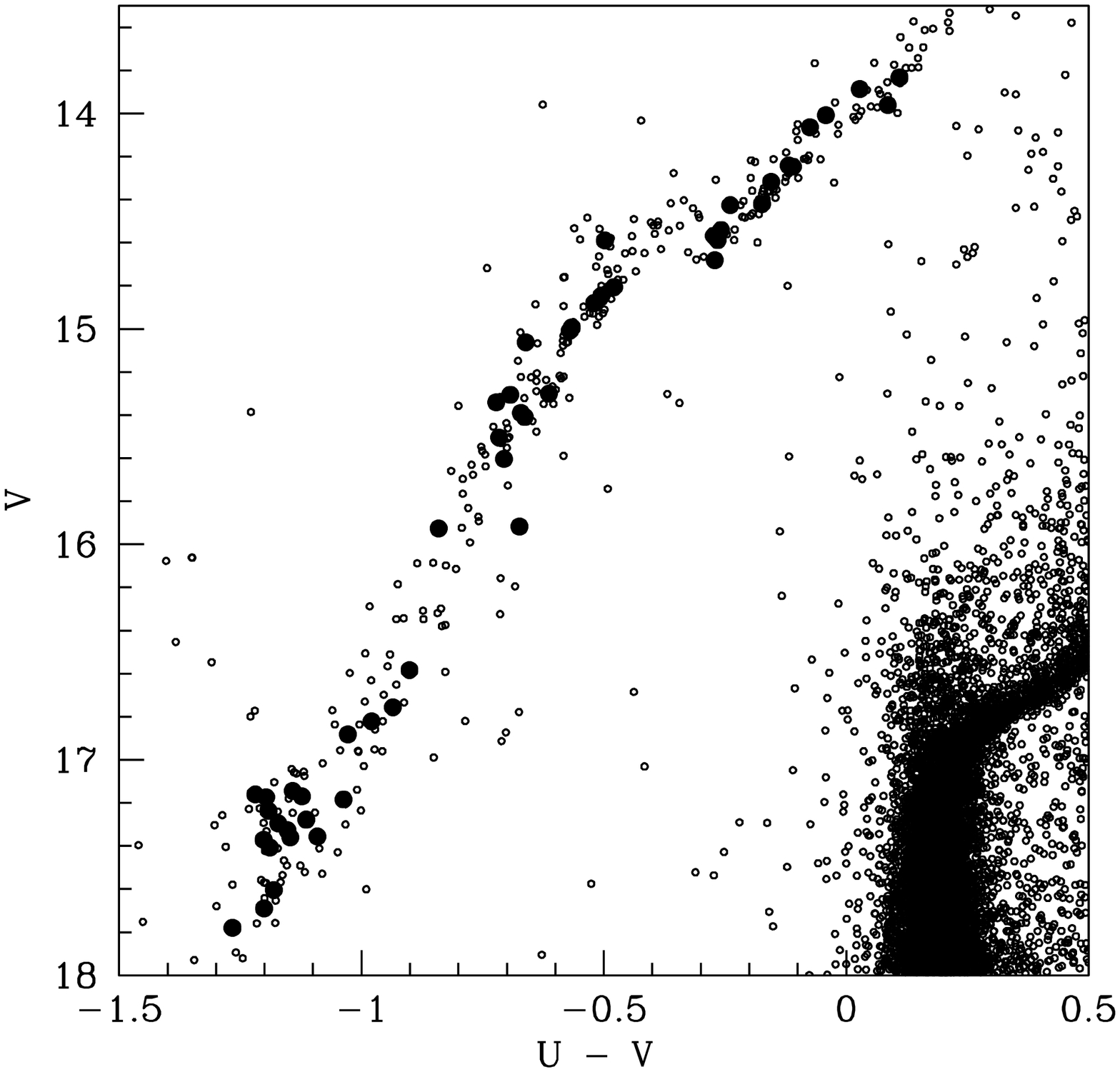,width=60mm,angle=0,clip=}}
\vskip1mm
\captionb{1}{Position of the observed stars in the color-magnitude diagram
of NGC\,6752. Data from Momany et al.~(2002).}
\vskip2mm
\centerline{\psfig{figure=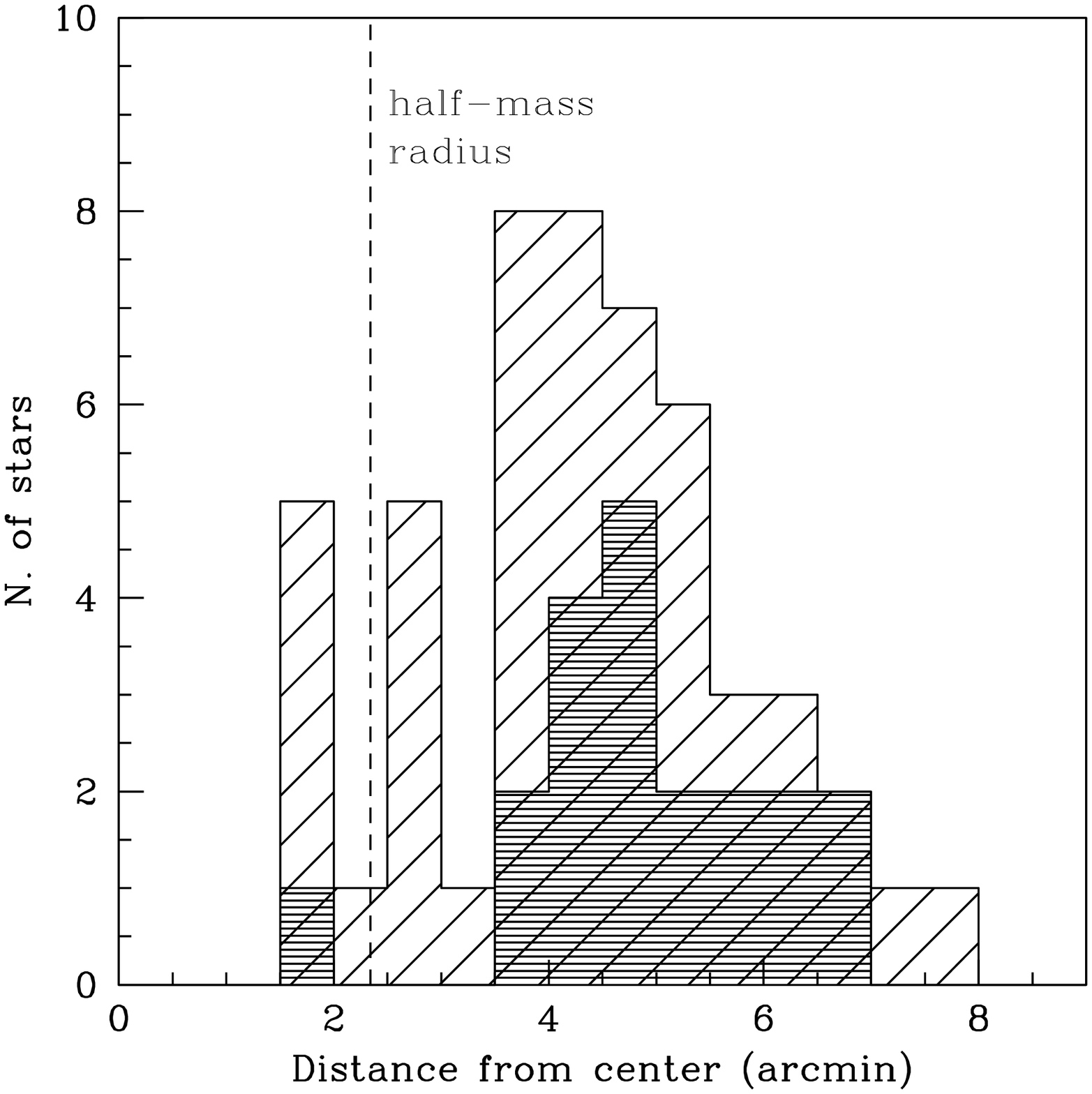,width=60mm,angle=0,clip=}}
\vskip1mm
\captionb{2}{Radial distribution of the observed stars. The dark shaded
area indicates hot stars (T$_{\mathrm{eff}}\geq20000$~K). The
half-mass radius from Harris (1996) is also indicated.}
}
\end{wrapfigure}

The data reduction has been performed with standard MIDAS procedures,
as described in detail in Moehler et al.~(2004) and Moni Bidin et al.
(2005).

During the same observing run similar data were acquired also for 32 HB
stars in M\,80.

\sectionb{3}{MEASUREMENTS}

Radial velocity (RV) variations have been measured with the cross-correlation
(CC) technique (Tonry \& Davis 1979) with the {\it fxcor} IRAF task. Before
coadding the single 1800s exposures in pairs we performed a CC between them
in order to verify that no significant RV variation had occurred.
For every star each spectrum has been cross-correlated with all the others,
performing 10 CCs for each star in field A and B, and 6 in field C, covering
different temporal intervals from one hour to 3.1 days.

Our analysis focused on the H$_{\beta}$ line, cross-correlating
the 4830-4890~\AA\ spectral range. Nevertheless, all the measurements
have been repeated also at other wavelengths, cross-correlating the
entire spectra, with and without H$_{\beta}$, and H$_{\gamma}$ when present
in the spectral range. In each CC the
position of the center of the cross-correlation function (CCF) has been determined
with a Gaussian fit (see for example Recio-Blanco et al.~2004,
for a description of the procedure). In the measurements on hot
stars it has been often impossible to cross-correlate the entire spectra without
the H$_{\beta}$ line, and we selected spectral intervals with the strongest
He lines instead, in order to minimise the noise included in the CC procedure
and obtain a CCF with a clear peak. We also applied a Fourier filter of various
shape (Brault \& White 1971)
to the noisy spectra, obtaining always better CCFs but substanitally
unchanged results.

The [OI] 5577~\AA\ sky line has been used as zero-point in order to correct the
spectral shifts due to differences between lamp and star spectra.

We have been forced to correct the data
for a systematic effect due to different positions of
the stars inside the slits
in different nights. The effect was up to
10-12~km~s$^{-1}$. On the slit images (without grism) acquired just
before each pair of exposures we measured the position of the stars
with respect to the center of the slit with a Gaussian fit of the stellar
profile parallel to the dispersion direction. Then we translated the
displacements between the frames to km~s$^{-1}$ with the instrumental
relation 1~pixel~=~38.2~km~s$^{-1}$, and applied them as corrections
to the RV variations.
A trend with Y position was evident,
but with a certain scatter due to random errors,
and we opted to derive the final corrections from the values
obtained from the least-square solution of this relation, in order to avoid to
introduce additional noise to the results.\\
This procedure gave corrections very similar to the RV variations
measured, indicating both that the removal of the systematic effect
had been succesful and that the RV variations observed were due only
to it.

\subsectionb{3.1}{Absolute RV measures}

We measured absolute RVs in order to check the cluster membership of
our targets, by means of CCs with the template star HD\,188112, a binary
sdB star with known ephemeris (Heber et al.~2003). These RVs have
undergone similar correction procedures as descibed before. The
errors (1$\sigma$) of these measures are 6-10~km~s$^{-1}$. All the
stars show an absolute RV in agreement with that of the cluster
($-$27.9~km~s$^{-1}$, Harris 1996) within 2$\sigma$, and can be
considered RV cluster members.

\sectionb{4}{ERROR ANALYSIS}
\begin{wrapfigure}[15]{l}[0pt]{5.5cm}
\vskip-2mm
\vbox{
\tabcolsep=4pt
\begin{tabular}{c|c}
\multicolumn{2}{c}{\parbox{5cm}{
~~~~{\bf Table 2.}{\
Errors estimates.}}}\\
\tablerule
error & range of values (km~s$^{-1}$) \\
\tablerule
\tablerule
$\sigma_{\mathrm{CC}}$ & 0.5-2 (H$_{\beta}$)\\
 & 1-5 (weak lines) \\
\tablerule
$\sigma_{\mathrm{wlc}}$ & 1.4-1.6 \\
\tablerule
$\sigma_{\mathrm{sky}}$ & 1.5 \\
\tablerule
$\sigma_{\mathrm{disp}}$ & 0.7-1.4 \\
\tablerule
$\sigma_{\mathrm{fit+ext}}$ &  0.7-1.5 (cooler stars) \\
 & 2.1-3.5 (hotter stars) \\
\tablerule
\end{tabular}
}
\end{wrapfigure}
The detection of binary systems in our survey is strongly dependent
upon a proper
estimate of the error budget. We performed  an accurate analysis of  all the
sources of errors and
estimated their values ($\sigma$).
Finally, all the error sources have been combined in quadrature.
The resulting  errors are  about
3-5~km~s$^{-1}$, with the exception of some measurements for hotter stars,
where the low S/N sometimes increased
the total error up to 7.5~km~s$^{-1}$.
In Table 2 we summerize
the contribution of each error sources and the corresponding ranges
($1\sigma$).
The CC error is evaluated directly from the CC theory (Tonry \& Davis 1979).\\
The
wavelength calibration (wlc) error has been measured on the lamp images calibrated
with the coefficients obtained in the wlc procedure, analyzing the position of nine
bright lamp lines.\\
The errors introduced by
the corrections for the sky line
position, $\sigma_{\mathrm{sky}}$, and for the displacement of the star inside the slits,
$\sigma_{\mathrm{disp}}$, have been estimated from the dispersion of the corrections
around the least-square solution when plotted against the Y-positions. \\
We identified two additional sources of errors: the choice of
the ``best fit'' of CCF peak, and the extraction of the spectra.  In
fact a different extraction can cause a slightly different line
profile; the CC procedure is sensitive enough to reveal
it. In measurements on hot stars these have been identified as the main sources of error.
We evaluated them together extracting all the spectra a second time in a
different manner and performing new measures in H$_{\beta}$ with
different fits, and, finally, measuring the dispersion $\sigma_{\mathrm{fit+ext}}$ of the
differences between these new data and the previous ones. This error was strongly dependent
on the S/N of the spectra, then we divided the targets into two groups
(stars cooler and hotter than $\approx$20000 K).

\sectionb{5}{RESULTS}

In Figure 3 we summarize the results obtained from measurements in
the H$_{\beta}$ wavelength range, where we obtained the most
reliable data. All the RV variations are small, lower than
15~km~s$^{-1}$, except for one star, and never greater than the
estimated 3$\sigma$ interval. They are only slightly higher for
hotter stars, but with larger errors due to decreasing S/N in the
spectra. We therefore conclude that {\it none of the observed RV
variations can be considered statistically significant}.

Star 15 in field A exhibited higher RV variations in in the third night,
and the higher datum plotted in the figure (21.7~km~s$^{-1}$) refers
to these measurements.  We consider this result interesting but
particularly dubious, because it was obtained in the only not-summed spectrum
(a single 1800s exposure, as mentioned before), where all the stars
tended to show higher RV variations due to the increased noise, and a
comparable variation is never seen in any of the other observing
nights.

\subsectionb{5.2}{Results from other lines}

The results from CCs involving the entire spectral range did not provide
additional information, because they always repeated the
results already obtained with H$_{\beta}$.  This is due to the
sensitivity of CC technique to the stronger line, and the extreme
difference in strength between Balmer and others lines in our spectra.

The results obtained with H$_{\gamma}$, when present in the spectra,
always confirm the ones obtained in H$_{\beta}$ wavelength range.
The differences are
always lower than 10~km~s$^{-1}$, with just a handful of exceptions in
agreement with a Gaussian distribution given the evaluated $\sigma$
value.

None of the variations measured with weak lines ever exceeds
the evaluated 3$\sigma$ value, confirming the previous conclusions.
The measurements with weak metallic lines have given very good results for
cooler stars, mainly in the range
11500~$\leq$~T$_{\mathrm{eff}}~\leq$~18000~K due to the presence of many
lines induced by radiative levitation of heavy elements (Glaspey et
al. 1989; Behr 2003).
In these cases the differences between the measured variations and the
ones in H$_{\beta}$ are really small, always below 4~km~s$^{-1}$.  For
hotter stars the low S/N and the lack of lines complicated the
measurements, and the results are less reliable. Occasionally the
difference reaches 30~km~s$^{-1}$, but always in the sense of limiting
the highest RV variations measured in H$_{\beta}$ and never
emphasizing them. In fact, the stars that shows the highest variations
in Figure 3 show no great variation in these measurements.

\vbox{
\centerline{\psfig{figure=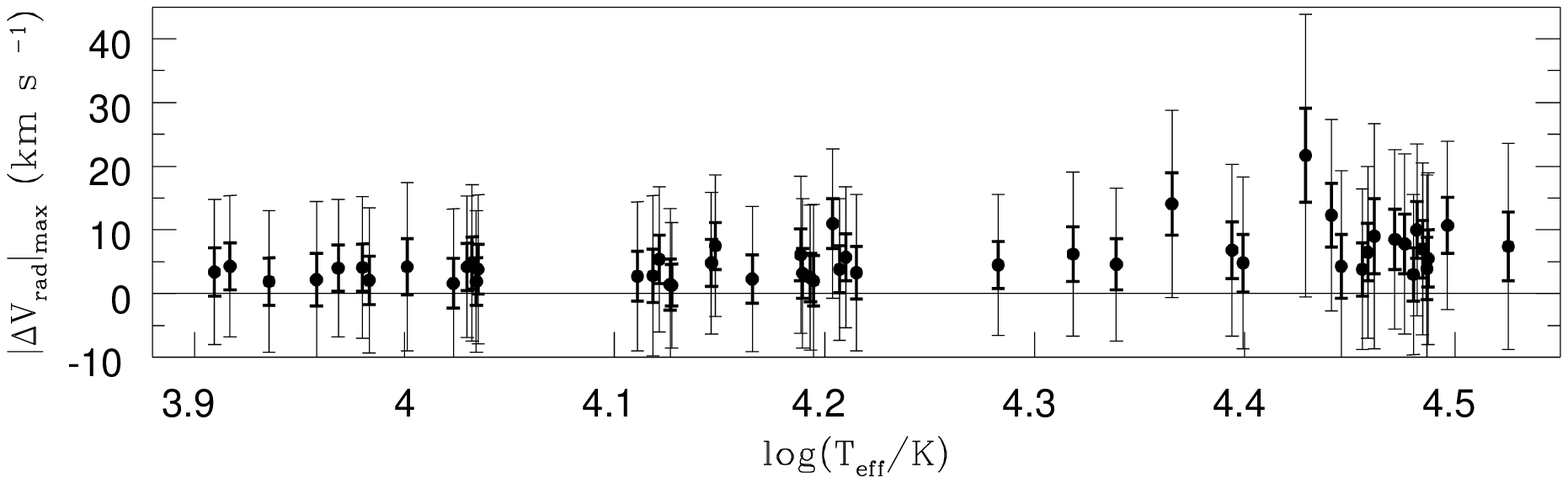,width=120truemm,angle=0,clip=}}
\captionb{3}{Maxima RV variations (in absolute value) measured in H$_{\beta}$
as a function of the temperature of the stars. The thick error bar indicates the
1$\sigma$ interval, meanwhile the thin one the 3$\sigma$.}
}
\vskip5mm

\begin{wrapfigure}[19]{r}[0pt]{62mm}
\vskip-2mm
\vbox{
\centerline{\psfig{figure=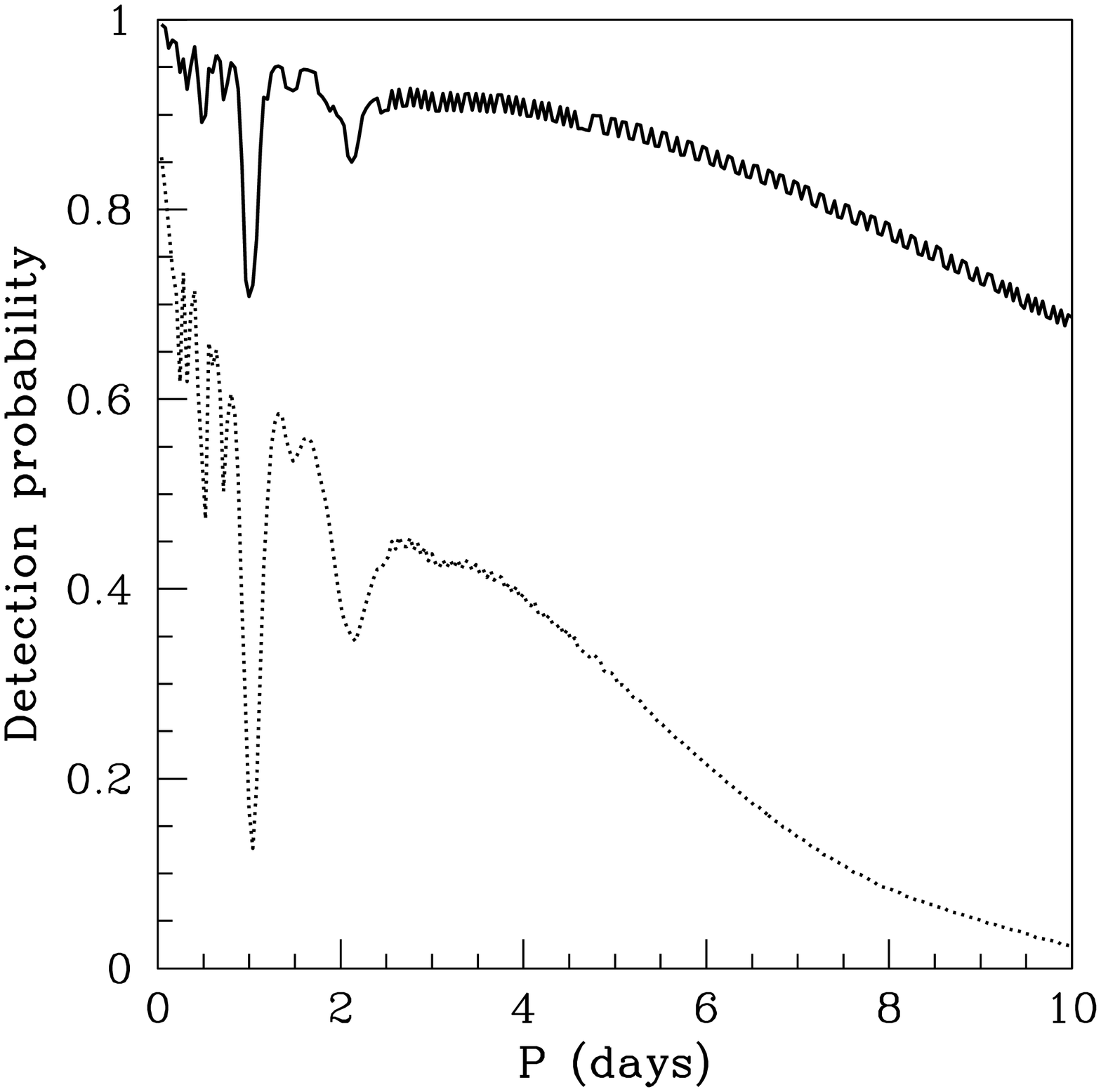,width=60mm,angle=0,clip=}}
\vskip1mm
\captionb{4}{Probability of detecting a binary in our observations.}
}
\end{wrapfigure}

\subsectionb{5.3}{Detection Probability}

In order to better understand our results we calculated the
probability of detecting a binary in our observations as a
function of the period P. We assumed circular orbits and a mass of
0.5~\msun\ for each component, in order to relate the semiamplitude of
the RV variation to the period. These assumptions are representative
of typical binary systems observed among field sdBs. 
We repeated the calculations assuming a companion of 0.1~\msun.
Systems with a low mass companion are a minority among field sdBs, but their
presence is well established. Our survey is not enough sensitive for
this kind of systems, except for the shortest periods, and from our results
we can't draw interesting conclusions about them.
The probabilities are shown in Figure 4 (solid line for a 0.5~\msun\ companion
and dotted line for 01~\msun).

\subsectionb{5.4}{Results for M\,80}

Since this is a work in progress we are still unable to show a plot
similar to Figure 3 for the stars observed in M\,80. Nevertheless we can
point out that the preliminary analysis of the first 15 stars in this cluster
gives results similar to
what obtained in NGC\,6752.
The variations are small,
being the highest values around 20~km~s$^{-1}$. Only two RV variations
exceed 15~km~s$^{-1}$, similarly to what was observed in NGC\,6752. It must
be emphasized that the results on M\,80 are going to be
less significant than in the case of NGC\,6752
both because the stars are much fainter and
the errors are larger, and because strong wind from north during
the observing run prevented us from observing this cluster in the
first two nights, so that the detection probability is reduced. We
evaluated a loss of sensitivity of about 20-25\% for periods up
to 5 days (compared to that on NGC\,6752), which, for longer periods,
drops rapidly to very low values.

\sectionb{6}{CONCLUSIONS}

The results shown in Figure 3 indicate that there is no close binary
system in our sample, neither among cooler stars, nor among EHB targets.
All the RV variations are within the estimated
errors, and are significantly lower than what expected. In the
compilation of Morales-Rueda et al.~(2003) for all field binary sdB
stars with known periods, the RV semi-amplitudes are always greater
than 30~km~s$^{-1}$. These binary systems would be easily detectable
in our survey. If the binary fraction of EHB stars in NGC\,6752 were
69\% as found by Maxted et al.~(2001) in the field, we would expect
that 13$\pm$2 of our 18 stars with T$_{\mathrm{eff}}\geq$22000~K
should be binaries. According to the period distribution found by
Morales-Rueda et al.~(2003) 80\% of them should have P$\leq$5 days,
and thus 9$\pm$2 binary systems should have been detected in our
sample. The absence of evidence of their presence is in sharp
contrast with these previous results, and can indicate a
significant difference between sdB stars in and outside globular
clusters.

In conclusion, most of the EHB stars in NGC\,6752, and possibly in M\,80, are
not close binaries. These results pose a number of problems. First
of all, we are forced to conclude that the dominant mechanism for the
formation of these stars in globular clusters does not involve the
interactions within a close binary system. Although the high close binary fraction
among field sdBs indicates that binary interactions can play an important role,
our results on the EHB stars in NGC 6752 indicates that there must be
other mechanisms at work.

It is possible that in GCs there are different formation channels of
EHB stars with respect to the main ones in the field.
For example, dynamical evolution of GCs could remove the primordial
binaries able to produce sdB stars, at least in the inner part of the
cluster. This is an interesting possibility, which has already been
proposed also to explain why the frequency of blue stragglers (BSs) in
GCs is significantly smaller than the frequency of field BSs (Piotto
et al.~2004), and the anticorrelation between the frequency of BSs in
GC cores and the GC total mass (Davies et al.~2004).  At the same
time, the GC environment can favour different formation mechanisms for
the EHB stars, like close encounters and collisions among stars (at
least in the cluster cores), or even more complex scenarios like the
ones proposed to explain the anomalous double main sequence population
in Omega Centauri (Bedin et al.~2004, and Piotto et al.~2005) and in
NGC\,2808 (D'Antona et al.~2005). Apparently, in these two clusters, a
second generation of stars could have formed from material polluted by
SNe and/or intermediate mass AGB star ejecta. These stars have an
enhanced He content, and could explain the presence of the EHB in both
clusters.

An extended survey of the presence of close binaries among cluster EHB
stars, from the inner core to the outskirts of NGC\,6752,
and of the presence of binaries with low companions or with
longer (100 days or so) periods in NGC\,6752 and other
clusters with EHBs, is absolutely needed in order to test the hypothesis of the
environment effects on the main production channels of these stars.

We note also that Peterson et al.~(2002) reported a high fraction of binaries
among the same type of stars in the same cluster, but Peterson (priv. comm.)
also pointed out that their sample is mainly located in the
outer regions of the cluster.

In concluding this brief discussion, we note that Napiwotzki et al.~(2004)
suggested that the low binary fraction among field sdBs found in their survey could
be due to contamination of the sample by halo and thick disk stars,
absent in Maxted et al.~(2001) sample since it included only bright
(thin disk) stars.
On average, our cluster EHB stars
are expected to be older, and possibly more metal poor than the
Napiwotzki et al.~sample, and apparently, the fraction of binaries among
them is even smaller than among thick disk-halo sdBs, suggesting a
possible dependence on ages or metallicities or on a
combination of these parameters.

\vskip1mm

ACKNOWLEDGMENTS.

\goodbreak
CMB and RAM acknowledge support by the Chilean Centro de Astrof\'isica
FONDAP (No. 15010003). GP acknowledges support by MIUR, under
the program PRIN03
\References

\refb
Bedin L.R., Piotto G., Anderson J., et al.~2004, ApJ, 605, L125
\refb
Behr B.B. 2003, ApJS, 149, 67
\refb
Brault J.W., White O.R. 1971, A\&A, 13, 169
\refb
D'Antona F., Bellazzini M., Caloi V., et al.~2005, ApJ, in press (astro-ph/0505347)
\refb
Davies M.B., Piotto G., De Angeli, F. 2004, MNRAS, 349, 129
\refb
Faulkner J. 1966, ApJ, 144, 978
\refb
Glaspey J.W., Michaud G., Moffat A.F.J., Demers S. 1989, ApJ, 339, 926
\refb
Han Z., Podsiadlowski P., Maxted P.F.L., Marsh T.R., Ivanova N. 2002,
MNRAS, 336, 449
\refb
Han Z., Podsiadlowski P., Maxted P.F.L., Marsh T.R. 2003, MNRAS, 341, 669
\refb
Harris W. E. 1996, AJ, 112, 1487
\refb
Heber U., Edelmann H., Lisker T., Napiwotzki R. 2003, A\&A, 411, L477
\refb
Heber U., Moehler S., Napiwotzki R., Thejll P., Green E.M. 2002, A\&A, 383, 938
\refb
Hoyle F., Schwarzschild M. 1955, ApJS, 2, 1
\refb
Lisker T., Heber U., Napiwotzki R., et al.~2005, A\&A, 430, 223
\refb
Maxted P.F.L., Heber U., Marsh T.R., North R.C. 2001, MNRAS, 326, 1391
\refb
Mengel J.G., Norris J., Gross P.G. 1976, ApJ, 204, 488
\refb
Moehler S., Sweigart A.V., Landsman W.B., Hammer N.J., Dreizler S. 2004, A\&A, 415, 313
\refb
Momany Y., Piotto G., Recio-Blanco A., et al.~2002, ApJ, 576, L65
\refb
Moni Bidin C., Moehler S., Piotto G., et al.~2005, A\&A, submitted
\refb
Morales-Rueda L., Maxted P.F.L., Marsh T.R., North R.C., Heber U. 2003,
MNRAS, 338, 752
\refb
Napiwotzki R., Karl C. A., Lisker T., et al.~2004, Ap\&SS, 291, 321
\refb
Peterson R.C., Green E.M., Rood R.T., Crocker D.A., Kraft R.P. 2002, in ASP
Conf. Ser. 265: Omega Centauri, A Unique Window into Astrophisics, 255
\refb
Piotto G., De Angeli F., King I. R., et al.~2004, ApJ, 604, L109
\refb
Piotto G., Villanova S., Bedin L.R., et al.~2005, ApJ, 621, 777
\refb
Recio-Blanco A., Piotto G., Aparicio A., Renzini A. 2004, A\&A, 417, 597
\refb
Tonry J., Davis M. 1979, AJ, 84, 1511

\end{document}